\documentclass[twocolumn,twoside,slac_two]{revtex4}
\usepackage{graphicx}
\usepackage{fancyhdr}
\usepackage{graphics}
\usepackage{epstopdf}
\usepackage{textpos}
\begin{document}

\title{\centering Status of the KATRIN experiment with special emphasis on source-related issues}
\author{
\centering
\begin{center}
Michael Sturm for the KATRIN Collaboration
\end{center}}
\affiliation{\centering Karlsruhe Institute of Technology (KIT), Institute for Technical Physics, Tritium Laboratory Karlsruhe (ITEP-TLK)}
\begin{abstract}
The Karlsruhe Tritium Neutrino experiment  KATRIN will allow a model independent measurement of the neutrino mass scale with an expected sensitivity of 0.2 eV/c$^2$ (90\% C.L.) and so will help to clarify the role of neutrinos in the early universe. KATRIN investigates spectroscopically the electron spectrum from tritium $\beta$-decay $^3$H $\rightarrow$ $^3$He + e$^-$ + $\bar\nu_e$ close to the kinematic endpoint of 18.6~keV with a high resolution electro-static filter of unprecedented energy resolution of $\Delta$E = 0.93 eV \cite{designrep}. KATRIN will be built at the Tritium Laboratory Karlsruhe on site of the KIT Campus North.

%
%
\end{abstract}

\maketitle
\thispagestyle{fancy}


\section{Introduction}
The properties of neutrinos and especially their rest mass play an important role at the intersections of cosmology,
particle physics and astroparticle physics. In standard cosmological models,
our universe is filled with primordial neutrinos arising
from freeze-out in the early universe. These neutrinos are
natural candidates for non-baryonic hot dark matter \cite{hanne}. 

A model-independent approach to determine
the neutrino mass is the kinematical analysis of electrons
from radioactive $\beta$- decay near the endpoint energy
E$_0$ \cite{otten}. A non-vanishing neutrino mass reduces the electron
endpoint energy and distorts the shape of the electron
spectrum.

%

\section{The KATRIN experiment}
\begin{figure*}[t]
\begin{center}
\includegraphics[width=165mm]{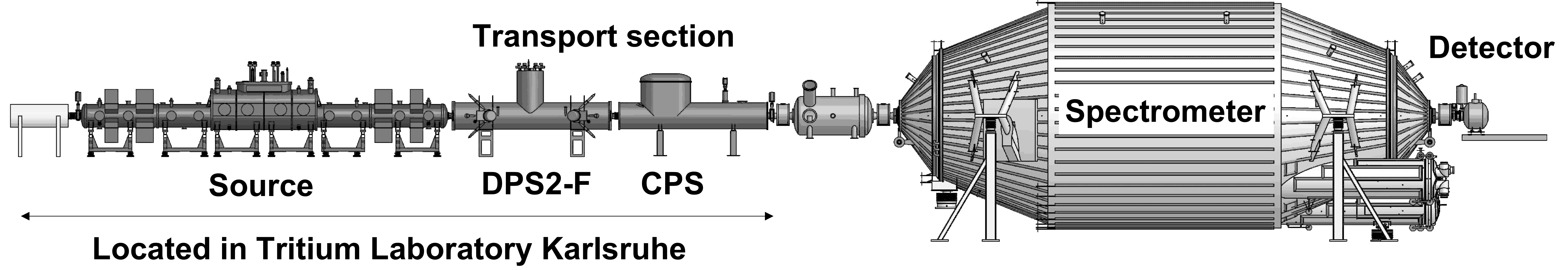}
\end{center}
\caption{Overview of the KATRIN main beam-line. High purity T$_2$ gas is being injected in the middle of the source tube with a flow rate of 1.8 mbar l/s (40 g(T$_2$)/day). Electrons from $\beta$-decay leave the source and are guided by magnetic fields through the transport section, while the remaining gas is being removed by active and cryogenic pumping. The pre-spectrometer filters out the low-energy part of the spectrum, thus only electrons close to the endpoint region can enter the main spectrometer for the precise energy analysis. Transmitted electrons are then detected by a low background Si-PIN detector system.} \label{fig:katrin}
\end{figure*}
A scheme of the 70~m long KATRIN setup is shown in figure \ref{fig:katrin}. KATRIN uses a strong windowless gaseous tritium source of almost pure molecular tritium (95\%) with a throughput of 40 g tritium per day stabilized on 0.1\% level. The decay electrons are guided adiabatically from the source through a transport section  to the spectrometer system by means of superconducting magnets while at the same time the tritium flow rate to the spectrometers has to be reduced by a factor  $> 10^{14}$, since the background rate generated by tritium decay within the spectrometers has to be less than 10$^{-3}$ counts/s in order to reach the sensitivity KATRIN is aiming for. The transport section consists of a differential pumping section (DPS2-F) and a cryogenic pumping section (CPS). In the DPS2-F the tritium flow will be reduced by differential pumping while in the CPS tritium will be adsorbed on a pre-condensed argon layer prepared inside KATRIN's beamtube.
In addition to the low background an high energy resolution as well as high statistics are a necessity. A tandem spectrometer system is used for energy analysis, followed by a detector-system for counting the transmitted $\beta$-decay electrons. Both spectrometers are of the MAC-E-Filter (Magnetic Adiabatic Collimation followed by Electrostatic Filter) \cite{mace}, \cite{mace2} type. 


The following subsections address the details of some key components related to the source and transport section and their present status.

\subsection{Windowless Gaseous Tritium Source}
\begin{figure}
\includegraphics[width=70mm]{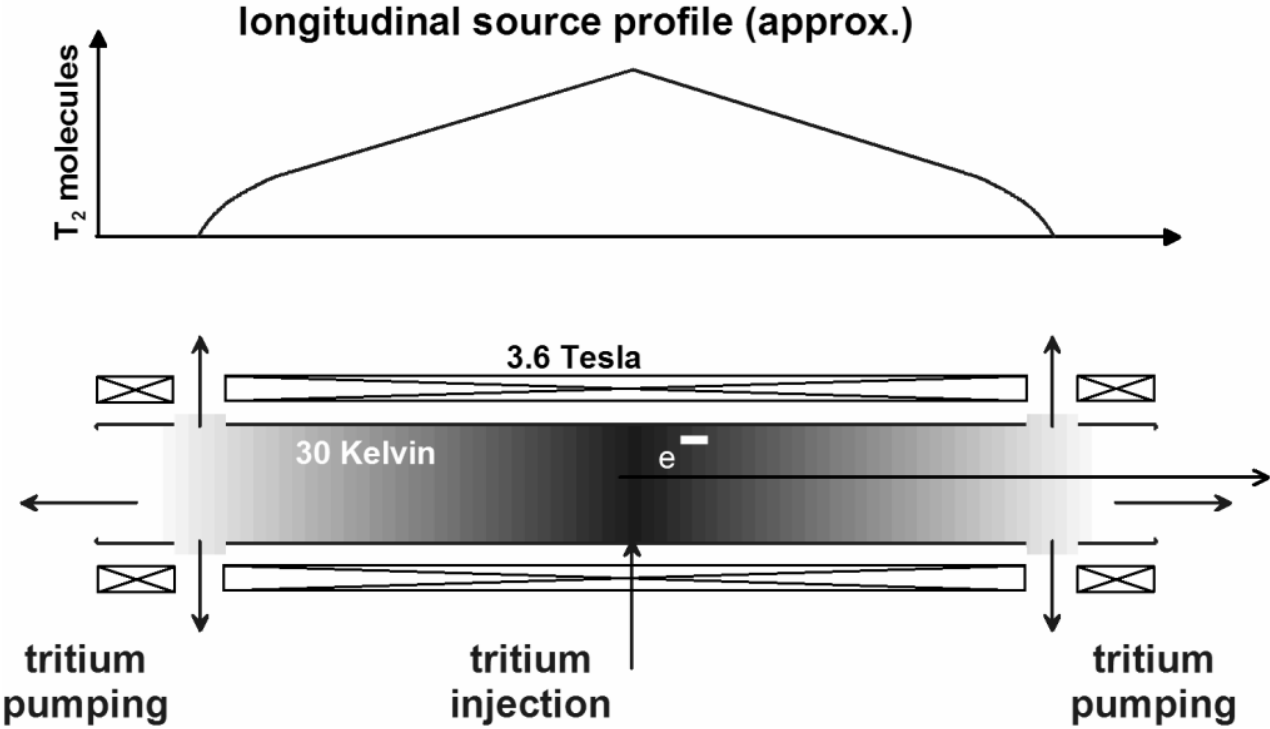}
\caption{Principle of KATRIN's Windowless Gaseous Tritium Source.}
\label{fig:wgts}
\end{figure}

One of the key parameters of KATRIN is the stability of the source on 0.1\% level.
 Figure \ref{fig:wgts} shows the principle of KATRIN's Windowless Gaseous Tritium Source (WGTS).
 At the center of the beamtube molecular tritium gas will be injected continuously.
After injection the T$_2$ molecules will diffuse to both ends of the WGTS beam-tube, where most of the tritium will be pumped out continuously  by turbomolecular pumps (TMP) in the first stages of the differential pumping section (DPS1), 
processed and reinjected. 
 The density profile of the gas inside the 10~m long beam-tube has to be kept stable on the 0.1\% level. Maintaining the required stable injection rate is  provided by the Inner Loop System discussed in the next section. In addition the temperature of the source beamtube has to be stabilized as well. Maintaining these conditions is
a very challenging task.

To stabilize the beam tube temperature with a stability of 0.1\% at a setpoint of about 30~K, two copper tubes are braced on the source beamtube which directly couple the beamtube to a two phase Neon thermosiphon \cite{demo}. This cooling concept is currently being tested at the WGTS demonstrator, a partly assembled version of the later WGTS cryostat which houses the later beam-tube and cryogenics but not the superconducting magnets. After finishing the demonstrator
tests the WGTS cryostat will be completed in order
to be ready for implementation into the KATRIN beam line.

With the first measurements a temperature stabilization in the milli-Kelvin range has been achieved, an improvement of a factor of 10 to 20 with regard to the specification.

\subsection{Inner Loop System}
\begin{figure}
\includegraphics[width=70mm]{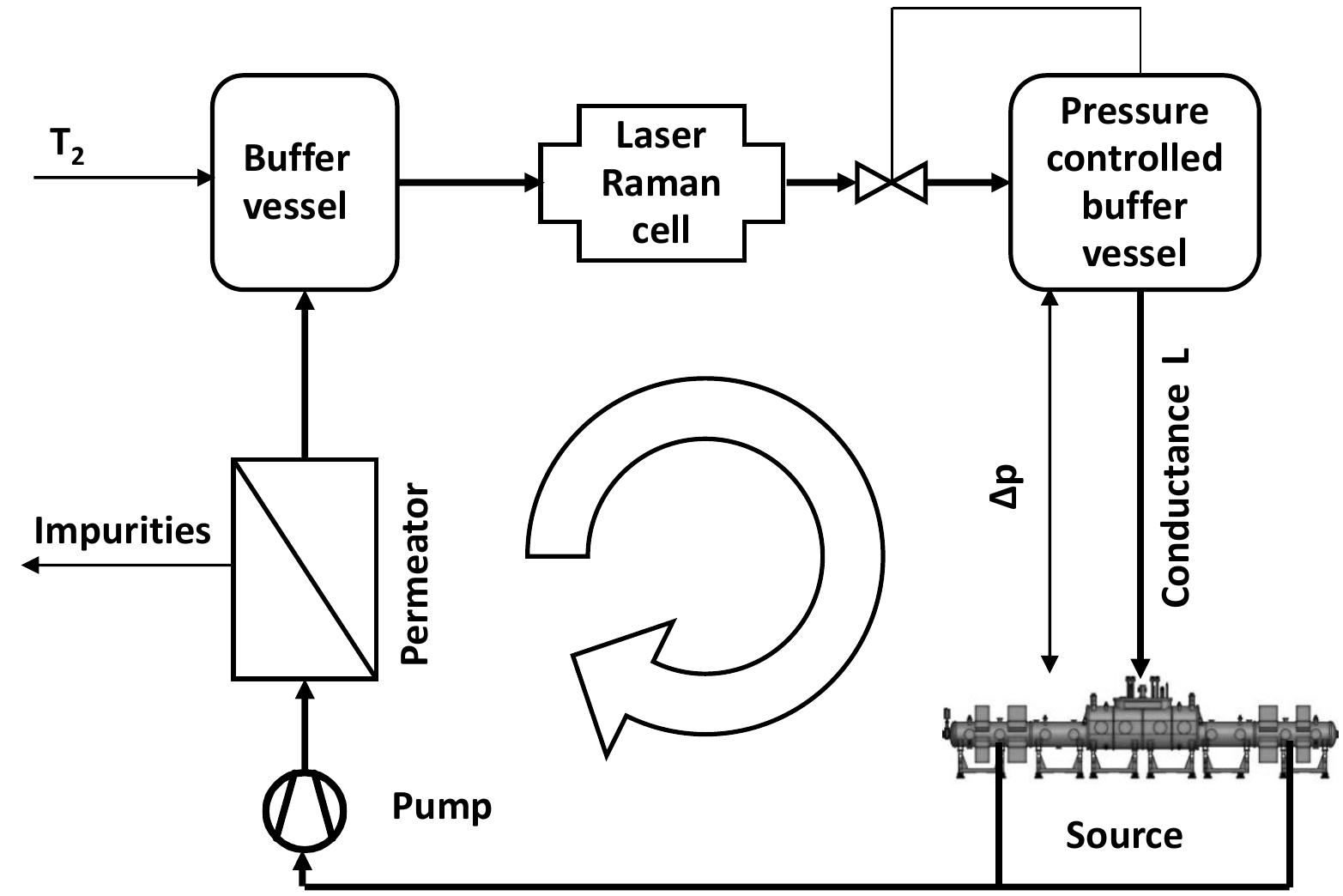}
\caption{Simpilfied flow diagram of the Inner Loop System.}
\label{fig:loop}
\end{figure}

In order to keep the pressure profile stable it is necessary to inject the tritium gas with a stability of 0.1\%, concerning flow-rate and composition. This challenging task is performed by the Inner Loop System, illustrated in a simplified flow diagram  in figure \ref{fig:loop}.
Tritium is being injected from a pressure controlled buffer vessel over a capillary with constant conductivity in the middle of the source beam tube.  The gas pumped out by the TMPs gets pumped through a palladium membrane filter (permeator) in a buffer vessel. From there the gas is led over a Laser Raman sampling cell and a regulating valve back into the pressure controlled buffer vessel. At the filter impurities like $^3$He from tritium decay and tritiated methanes, being generated due to interactions of tritum with the carbon inside the stainless steel walls of the system, are being detached from the gas stream. The amount of gas not recirculated is replaced by tritium from TLK's Isotope Separation System. 

The Inner Loop System has been set up and commissioned successfully. The first test-runs have been performed with a capillary of similar conductance as KATRIN's injection capillary and source tube \cite{loops}. 
Being designed for a stability of 10$^{-3}$, these test runs showed that the loop system reaches a $2\cdot 10^{-4}$ stability level during 4 week of continuous gas circulation.

\subsection{Laser Raman Spectroscopy}
Any small change of the tritium gas composition will manifest itself in non-negligible effects
on the KATRIN measurements; therefore, precise methods to specifically monitor the gas composition have to be
implemented. Laser Raman Spectroscopy is the method of choice for the monitoring of the gas composition because
it is a non-invasive and fast in-line measurement technique.
Laser Raman spectroscopy allows to monitor all hydrogen
isotopologues (T$_2$, DT, D$_2$, HT, HD, H$_2$) simultaneously \cite{lara1}. Before entering the injection vessel in the above described Inner Loop system, the gas passes a Laser Raman cell in which the gas is being analyzed.
 Measurements on flowing non-tritiated
\cite{lara2} and tritiated \cite{lara3} gas samples have been performed.
A level of detection of 3 mbar partial pressure in 1~s acquisition time
has been achieved \cite{lara4}.

\subsection{Differential Pumping Section}
The DPS2-F represents the last stage of turbomolecular pumping in the transport section. Its reduction of the gas-flow in direction of the spectrometers is of high importance to achieve the 14 orders of flow rate reduction from gas injection point in WGTS to the end of the transport section. At the DPS2-F the gas flow to the spectrometers will be reduced by 4 differential pumping TMPs by about 5 orders of magnitude.

Another item the DPS2-F has to deal with is that in the KATRIN source tube ions and ion clusters are being created. An accumulation of ions inside the beamtube would distort the $\beta$-spectrum, in addition these ions are not effectively pumped out, since they are confined by the magnetic field in a similar way as the electrons. To prevent this, a dipole system for ion removal as
well as two Fourier-transform-ion cyclotron resonance (FTICR) \cite{fticr}
ion traps for ion-concentration monitoring will be installed
in the beamtube of the DPS2-F. This instrumentation also increases the gas-flow reduction factor compared to the bare beamtube by reducing the inner beam-tube diameter \cite{dps2}.
Up to now measurements of the reduction factor with various gases have been performed without instrumentation inside the beamtube. The measured gas-flow reduction factors for this geometry are in good agreement with simulations \cite{dps}.
\section{Conclusion}
KATRIN has ambitious goals, both
in particle physics and in the technical realization of the
experimental set-up. Currently the spectrometers, the detector and the DPS2-F cryostat are on site.
The first measurements for the temperature stabilization of the source beam tube with the demonstrator, as well as the first gas-flow reduction factor measurements with the DPS2-F have been successful. The Inner Loop system has been successfully commissioned and Laser Raman Spectroscopy for monitoring the gas composition has been successfully implemented.
 The next major steps to finalize the experiment setup will be finishing the manufacturing of the WGTS cryostat and the delivery of the CPS. After delivery, installation test of all components the experiment will be ready for the first tritium measurements.
\begin{acknowledgments}
The work of the author was supported in part by BMBF (05A08PM1) and DFG SFB Transregio 27 "Neutrinos and Beyond".
\end{acknowledgments}
\bigskip 

\end{document}